\documentclass[twocolumn]{elsart}

\usepackage{amssymb}

\begin{document}

\begin{frontmatter}



\title{Vortex dynamics in dilute two dimensional Josephson junction arrays}


\author{Md. Ashrafuzzaman and Hans Beck}
\ead{md.ashrafuzzaman@unine.ch\\Hans.beck@unine.ch}

\address{Institute of Physics, University of Neuchatel, Switzerland}

\begin{abstract}
We have investigated the dynamics of vortices in a dilute two dimensional Josephson junction array where a fraction of the superconducting islands is
missing. We have used the multiple trapping model to calculate the mobility of vortices and the frequency dependence of the resistance and inductance of the array.
\end{abstract}

\end{frontmatter}

\label{}
We consider an infinite two dimensional (2D) Josephson junction array (JJA), where each site is randomly occupied by a superconducting island with some
probability $p$ ($p=1$ for regular array) and empty with probability $1-p$. At a  lower
critical  value of $p=p_c$ ($p_c=0.5$ for triangular array, for site percolation) the superconducting properties [1,2]
of the arrays are destroyed. This type of arrays undergoes a transition from superconducting to normal states at a critical temperature which is called
Berezinskii-Kosterlitz-Thouless transition temperature ($T_{BKT}$). Experimentally [3] the frequency dependence of the conductance $G=1/R_s+1/i\Omega L_s$ of
the sheet (array) is observed ($\Omega$ is the frequency). We calculate the contribution to the resistance
($R_s$) and inductance ($L_s$) produced by moving vortices (V) and antivortices (A). For this we have to first find the mobility $\mu$ of the V and A in the
disordered  array and for it we use the  multiple trapping model (MTM),
developed for electronic transport in amorphous semiconductors [4]. A vortex moves
randomly in the array and whenever it reaches a hole (area of missing sites) it experiences some pinning potential by the hole and falls into it.
With some probability it can also be released from the hole due to thermal excitations. We have
considered the holes of missing sites in 2D, for simplicity, to be circular. In this model the vortex mobility $\mu$ is given by
$\mu(\omega)=\frac{\mu_0}{1+\int dN\frac{N}{pN_{t}}\frac{D(N)}{i\omega+e^{-\beta E(N)}}}\equiv (\nu'(\omega)+i\nu''(\omega))\mu_0$
where $\nu(\omega)=\frac{\mu(\omega)}{\mu_0}=\nu'(\omega)+i\nu''(\omega)$. $\mu_0$ is the mobility of V or A in a regular JJA.
$D(N)$ is the number density of holes with $N$ missing sites taken from some experimental fit [5] and $N_t$ is the total number of sites of the array.
$E(N)=\frac{1-p_c}{p-p_c}2k_BT_{BKT}\ln\sqrt{N+1}$ is the binding energy in a circular hole, $\beta=\frac{1}{k_BT}$, $T$ is the temperature.\\
In our model we get  the following expressions for $R_s$ and $L_s$
\begin{equation}
\frac{1}{R_s}=\frac{1}{R_0}+\frac{1}{L_0 \omega_a}\frac{A'}{(\frac{r}{\omega_a}\omega+A'')^2+A'^2}
\end{equation}
\begin{equation}
\frac{1}{L_s}=\frac{r}{L_0\omega_a}\frac{\omega(\omega+A'')}{(\frac{r}{\omega_a}\omega+A'')^2+A'^2}
\end{equation}
where $A'=8\pi\frac{1-p_c}{p-p_c}\frac{1}{t}\bar{n}\nu'(\omega)$ and $A''=8\pi\frac{1-p_c}{p-p_c}\frac{1}{t}\bar{n}\nu''(\omega)$, $t=\frac{T}{T_{BKT}}$, $\bar{n}$ is the
vortex density, $\omega=\Omega/\omega_a$ is the scaled frequency and $\omega_a(\approx 10^{6}Hz)$ is the Debye frequency of the lattice. $R_0$ and $L_0$ are the
resistance and inductance resulting from the currents flowing in the junction in absence of vortices.\\
\begin{figure} [h]
\let\picnaturalsize=N
\def\picsize{6.0 cm}
\def\picfilename{RsJMMcor.eps}
\ifx\nopictures Y\else{\ifx\epsfloaded Y\else\input epsf \fi
\let\epsfloaded=Y
\centerline{\ifx\picnaturalsize N\epsfxsize \picsize\fi \epsfbox{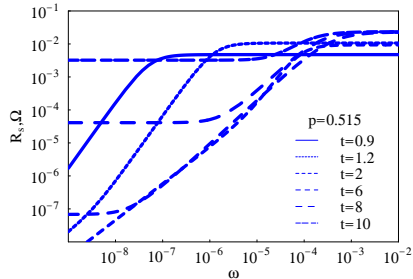}}}\fi
\caption{{\bf $R_s(\omega)$ vs $\omega$ for different $T$.}}
\label{AmplitudePhi4}
\end{figure}
\begin{figure} [h]
\let\picnaturalsize=N
\def\picsize{6.0 cm}
\def\picfilename{LsJMM.eps}
\ifx\nopictures Y\else{\ifx\epsfloaded Y\else\input epsf \fi
\let\epsfloaded=Y
\centerline{\ifx\picnaturalsize N\epsfxsize \picsize\fi \epsfbox{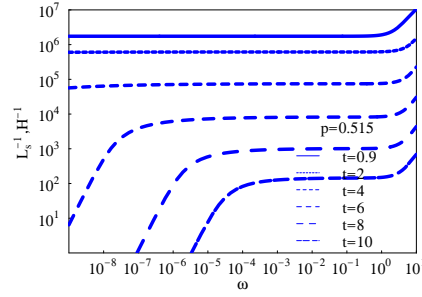}}}\fi
\caption{{\bf $L_s^{-1}(\omega)$ vs $\omega$  for different $T$.}}
\label{AmplitudePhi4}
\end{figure}
Three types of dependences of $R_s$ on $\omega$ are visible. At very low and high $\omega$ white spectrums are observed. Inbetween $R_s\propto\omega^x$
with $x=1$ for higher $p$ ($p\approx 1$) but for lower $p$ ($p\to p_c$), $x$ rises to 2, respectively 4/3, for $T=T_{BKT}$ respectively $T=10T_{BKT}$.
For inductive part $L_s^{-1}\propto\omega^2$ at low $\omega$ while at high $\omega$, $L_s^{-1}\propto\omega$ leaving an intermediate frequency range where
$L_s^{-1}$ is independent of $\omega$. The critical frequency for
crossover from $L_s^{-1}\propto\omega^2$ to constant $L_s^{-1}$ decreases with the decrease of $T$ and increase of $p$.  Our results are in good agreement with
the experimental data [3].\\
We thank P Martinoli for valuable discussions. The Swiss National Science Foundation supported this work under the project no 2000-067853.02/1.\\
References\\
\begin{scriptsize}
$[1]$ R.S. Newrock, C.J.Lobb,U.Geigenm$\ddot{u}$ller, M.Octavio ; Solid State Physics 54, 263(2000)\\
$[2]$ Md. Ashrafuzzaman and Hans Beck, Studies of High Temperature Superconductors, Nova Science Pub,NY,2002,V43,ch5, also cond-mat/0207572\\
$[3]$ J$\acute{e}$r$\hat{o}$me Affolter and P. Martinoli, Institute of Physics, University of Neuchatel, Switzerland (2001), Unpublished\\
$[4]$ H. Scher et al., Phys Rev B12, 2455 (1975)\\
$[5]$ C. Bailat and H. Beck, Institute of Physics, University of Neuchatel, Switzerland, unpublished\\
\end{scriptsize}










\end{document}